\documentclass[aps,prl,twocolumn,superscriptaddress,reprint,showpacs]{revtex4-2}
\usepackage{graphicx}
\usepackage{setspace}
\usepackage{amsmath}
\usepackage{amssymb}
\usepackage{color}
\usepackage{array}
\usepackage{subfigure}
\usepackage{hyperref}
\usepackage{float}
\usepackage{lipsum}

\usepackage[all]{xy}
\newcommand{\RN}[1]{%
  \textup{\uppercase\expandafter{\romannumeral#1}}%
}

\begin{document}
\title{Analysis of heralded higher-fidelity two-qubit entangling gates with self-correction}

\author{Yuan Sun}
\email[email: ]{yuansun@siom.ac.cn}
\affiliation{CAS Key Laboratory of Quantum Optics and Center of Cold Atom Physics, Shanghai Institute of Optics and Fine Mechanics, Chinese Academy of Sciences, Shanghai 201800, China}
\affiliation{University of Chinese Academy of Sciences, Beijing 100049, China}

\begin{abstract}
For the quantum error correction (QEC) and noisy intermediate-scale quantum (NISQ) algorithms to function with high efficiency, the raw fidelity of quantum logic gates on physical qubits needs to satisfy strict requirement. The neutral atom quantum computing equipped with Rydberg blockade gates has made impressive progress recently, which makes it worthwhile to explore its potential in the two-qubit entangling gates, including Controlled-PHASE gate and in particular the CZ gate. Provided the quantum coherence is well preserved, improving the fidelity of Rydberg blockade gates calls for special mechanisms to deal with adverse effects caused by realistic experimental conditions. Here the heralded very-high-fidelity Rydberg blockade Controlled-PHASE gate is designed to address these issues, which contains self-correction and projection as the key steps. This trailblazing method can be built on the basis of the previously established buffer-atom-mediated gate, and a special form of symmetry under PT transformation plays a crucial role in the process. We further analyze the performance with respect to a few typical sources of imperfections. This procedure can also be regarded as quantum hardware error correction or mitigation. While this paper by itself does not cover every single subtle issue and still contains many over-simplifications, we find it reasonable to anticipate very-high-fidelity two-qubit quantum logic gate operated in the sense of heralded but probabilistic, whose gate error can reduce to the level of $10^{-4}$--$10^{-6}$ or even lower with reasonably high possibilities.
\end{abstract}
\maketitle

The research of quantum computing has now reached the era of noisy intermediate-scale quantum (NISQ) devices and the power of quantum error correction (QEC) has emerged on the horizon \cite{GoogleRQCnature2024, Google2024Willow}. For practically running quantum computing tasks with high efficiency to surpass supercomputer \cite{Wujunjie2018NSR} and for constructing logical qubits \cite{SurfaceCode2002, Kitaev20032, PhysRevA.86.032324}, it definitely requires the physical qubits to operate quantum logic gates with very high fidelity, in particular the two-qubit entangling gates. This problem naturally has close ties with the specific properties of physical qubits, and the quality of two-qubit gates on various physical platforms still has a relatively large gap towards the desired level at this moment. Among many competitive candidates, the cold atom qubit \cite{PhysRevA.62.052302, RevModPhys.82.2313, Saffman_NSR} seems particularly interesting and arouses a lot of attentions recently with the solid advancement in two-qubit gate fidelity \cite{PhysRevA.105.042430, Lukin2023nature} and long coherence time. Therefore neutral atom quantum computing becomes the attractive choice of the paradigm to explore the potential of two-qubit gates with the emphasis on obtaining much higher fidelity in experimental implementations.

Ever since the early Rydberg blockade two-qubit gate experiments \cite{nphys1178, PhysRevLett.104.010503}, the Rydberg dipole-dipole interaction \cite{PhysRevLett.102.013004} plays an essential role and such that coherently driving ground-Rydberg transition is indispensable. Together with the progress of laser technologies, the improvement of neutral atom two-qubit gate comes from using synthetic continuously-modulated pulse to replace the early proposal of discrete resonant pulse sequences \cite{PhysRevApplied.13.024059, PhysRevApplied.15.054020}. Provided the number of cold atom qubits can exceed $\sim 1000$ \cite{Pause2024Optica}, the proposed solution for high-speed high-connectivity large-scale array is the buffer atom framework, consisting of the buffer-atom-mediated (BAM) gate \cite{Yuan2024SCPMA, Yuan2024FR2} and Rydberg blockade SWAP gate \cite{Yuan2024SWAP}. Although these developments so far indicate a promising platform to instantiate quantum processors, nevertheless the two-qubit gate fidelity won't surpass the level of $\sim 0.999$ with optimistic estimation even with well-tuned post-selection techniques.

In this paper, we establish a trailblazing method to improve the fidelity of two-qubit Rydberg blockade gate as well as systematically analyzing its working principles and potential performances. It employs an atom-laser interaction process with an embedded symmetry under PT (Parity-Time) transformation, in which a self-correction mechanism can be arranged to almost cancel the deviations caused by various sources. It concludes with a projection stage as the heralding operation. On the algorithm side, this category of methods feel like hardware-level quantum error correction or mitigation, while on the physics side it feels like in-situ comparison of wave functions during the time evolution process. The rest of paper is organized as follows. At first, we introduce the basic mechanisms on the basis of previously established BAM gate and demonstrate the typical routines of design. Then we discuss the properties of this special heralded gate with respect to many typical adverse effects. We adhere to suppressing the high-frequency components in the waveforms \cite{PhysRevApplied.20.L061002} of relevant atom-laser interaction. The main contents here focus on the one-photon ground-Rydberg transition and are in fact readily applicable to two-photon \cite{OptEx480513} or even three-photon ground-Rydberg transitions. The two-qubit gate fidelity will be evaluated according to the well-established standard criteria \cite{PEDERSEN200747, PhysRevA.96.042306}.

Many adverse effects can cause the experimental conditions to deviate from the idealized gate operation, and the method here aims at offsetting the errors accumulated on the wave functions with a heralding signal to indicate success. The anticipated tactic will meet several prerequisites, including a reasonably small extra time consumption, a reasonably low extra laser power requirement and a minimal extra phase accumulation for the idealized case. 

\begin{figure}[h]
\centering
\includegraphics[width=0.46\textwidth]{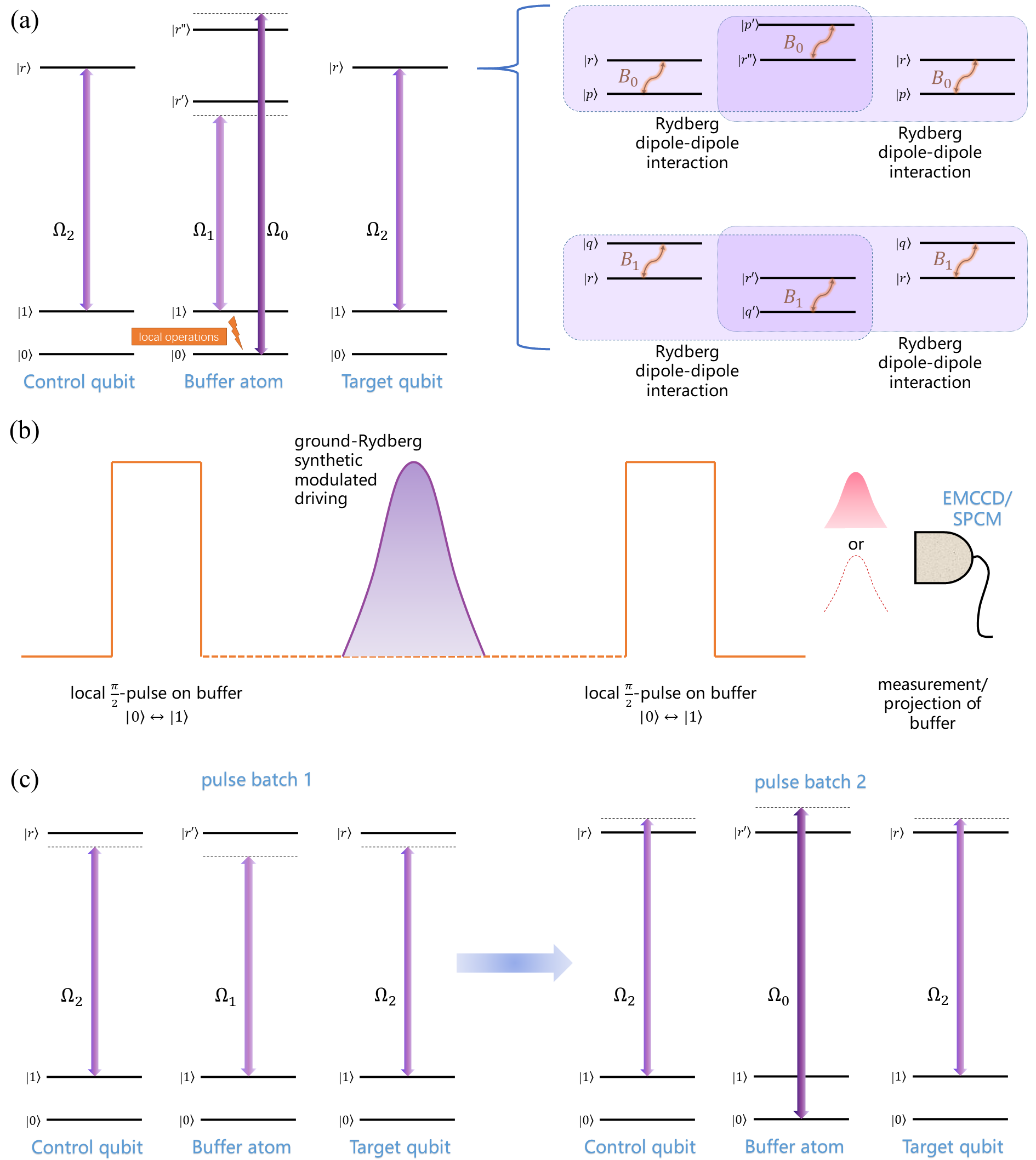}
\caption{(Color online) The (a) linkage structure and (b) basic pulse sequence of the atom-laser interaction process under study. The dual-rail BAM gates involves both $|0_\text{b}\rangle, |1_\text{b}\rangle$ and the necessary inter-atomic interaction can come from two Rydberg dipole-dipole channels as in (a) or only one Rydberg dipole-dipole channel but with different magnetic substates. The subscripts $\text{b, c, t}$ denote the buffer, control and target atoms respectively in the text. Particularly, in (a) the buffer atom receives off-resonant ground-Rydberg driving while the qubit atoms receive resonant driving \cite{Yuan2024FR2}. (c) shows the pulse sequence variation form of the dual-rail technique.}
\label{fig:layout_sketch}
\end{figure}

We carry out the formal derivations on the foundation of the previously established buffer-atom-mediated gate \cite{Yuan2024SCPMA, Yuan2024FR2} and we prefer to choose difference elements for the buffer and qubit atoms such that their optical drivings do not interfere. With the basic procedures shown in Fig. \ref{fig:layout_sketch}, the buffer atom now needs the qubit register states $|0\rangle, |1\rangle$. In order to enforce the dual-rail BAM gates, there exists two separate Rydberg dipole-dipole interaction channels $|r'\rangle\leftrightarrow|r\rangle, |r''\rangle\leftrightarrow|r\rangle$ modeled by the F\"orster resonance structure. It is also possible to utilize only one Rydberg level of the buffer atom, for example, by utilizing the polarization degree of freedom and letting $\Omega_0, \Omega_1$ couple to different magnetic substates of the same Rydberg level (see supplemental material for more details). We summarize the gate operations as the following three steps. (\RN{1}) Prepare the buffer atom in $|0_\text{b}\rangle$ and then apply a local $\frac{\pi}{2}$-pulse to render $\big(|0_\text{b}\rangle - i|1_\text{b}\rangle\big)/\sqrt{2}$. (\RN{2}) Apply the special symmetric form of composite buffer-atom-mediated gate involving both $|0_\text{b}\rangle, |1_\text{b}\rangle$ as the buffer states. More specially, this feels like the linear superposition of two buffer-atom-mediated gates as PT-reversal of each other and we will see the necessity of this later. (\RN{3}) Apply a second local $\frac{\pi}{2}$-pulse on the buffer atom and measure its projection on its local basis $|0_\text{b}\rangle, |1_\text{b}\rangle$. 

Next we continue to explain the heralding and self-correction mechanisms. Without loss of generality, let's postulate that the purpose is to implement CZ gate on some two-qubit system as $C_{00}|00\rangle + C_{01}|01\rangle + C_{10}|10\rangle + C_{11}|11\rangle$ $\to$ $C_{00}|00\rangle + C_{01}|01\rangle + C_{10}|10\rangle - C_{11}|11\rangle$ up to an overall phase factor. After step \RN{1}, the buffer-qubit system reaches the state of $\big(|0_\text{b}\rangle - i|1_\text{b}\rangle\big)\big( C_{00}|00\rangle + C_{01}|01\rangle + C_{10}|10\rangle + C_{11}|11\rangle\big)/\sqrt{2}$. The consequence of step \RN{2} amounts to two simultaneous BAM CZ gates: one with $|0_b\rangle, \Omega_0$ and the other one with $|1_b\rangle, \Omega_1$. This results in $\big(|0_\text{b}\rangle - ie^{i\eta}|1_\text{b}\rangle\big)\big( C_{00}|00\rangle + C_{01}|01\rangle + C_{10}|10\rangle - C_{11}|11\rangle\big)/\sqrt{2}$ and $e^{i\eta}$ marks the relative overall phase difference between the two BAM CZ gates which can be corrected for locally on the buffer atom. So far in the idealized case it does not look like any more superior than the original BAM gate. But, now consider some experimental adverse effects occurring, and henceforth the dual-rail BAM gates receive deviations correspondingly: $|0_\text{b}\rangle \big( (1+\delta u_0)C_{00}|00\rangle + (1+\delta v_0)C_{01}|01\rangle + (1+ \delta w_0)C_{10}|10\rangle - (1+\delta z_0)C_{11}|11\rangle\big)/\sqrt{2} - i|1_\text{b}\rangle \big( (1+\delta u_1)C_{00}|00\rangle + (1+\delta v_1)C_{01}|01\rangle + (1+ \delta w_1)C_{10}|10\rangle - (1+\delta z_1)C_{11}|11\rangle\big)/\sqrt{2}$. The appropriate second local $\frac{\pi}{2}$-pulse will generate the state of $|0_\text{b}\rangle \big( (2+\delta u_0 + \delta u_1)C_{00}|00\rangle + (2+\delta v_0 + \delta v_1)C_{01}|01\rangle + (2+\delta w_0 + \delta w_1)C_{10}|10\rangle - (2+\delta z_0 + \delta z_1)C_{11}|11\rangle \big)/2+i|1_\text{b}\rangle \big( (\delta u_0 - \delta u_1)C_{00}|00\rangle + (\delta v_0 - \delta v_1)C_{01}|01\rangle + (\delta w_0 - \delta w_1)C_{10}|10\rangle - (\delta z_0 - \delta z_1)C_{11}|11\rangle\big)/2$.

Currently, the experimental techniques of quantum control can suppress the two-qubit gate error to the level of $\sim 0.001$ on various platforms including the neutral atom, ion and superconducting qubits. In other words, it makes sense to treat $\delta u_0, \delta v_0, \delta w_0, \delta z_0, \delta u_1, \delta v_1, \delta w_1, \delta z_1$ as relatively small perturbations. If we can design the synthetic drivings to satisfy the following condition:
\begin{eqnarray}
\delta u_0 = -\delta u_1 \equiv \delta u, \delta v_0 = -\delta v_1 \equiv \delta v, \nonumber\\
\delta w_0 = -\delta w_1 \equiv \delta w, \delta z_0 = -\delta z_1 \equiv \delta z,
\label{eq:waveform_requirement}
\end{eqnarray}
where these relations may hold strictly or at least valid up to the 1st order with respect to the specific perturbation, and then the state of the system eventually becomes: $|0_\text{b}\rangle\big(C_{00}|00\rangle + C_{01}|01\rangle + C_{10}|10\rangle - C_{11}|11\rangle\big) + i|1_\text{b}\rangle\big( \delta u C_{00}|00\rangle + \delta v C_{01}|01\rangle + \delta w C_{10}|10\rangle - \delta z C_{11}|11\rangle \big)$. This separation of wave functions into the two parts in the forms of $\delta x_0 + \delta x_1$ and $\delta x_0 - \delta x_1$ essentially corresponds to unitarity. For the first-order perturbations caused by various adverse effects, the responses seem to be first-order phase deviations of wave functions in Rydberg gates via synthetic modulated drivings \cite{PhysRevApplied.13.024059, OptEx480513}. After projection, the outcome of $|0_\text{b}\rangle$ indicates the success of the higher fidelity gate while $|1_\text{b}\rangle$ indicates gate error. Throughout the text the gate error is defined as $1-f$ with $f$ representing the fidelity.

The path forward clearly relies on finding the dual-rail waveforms satisfying the requirement of Eq. \eqref{eq:waveform_requirement}. This certainly has close relations with the physical sources and properties of the specific adverse effect under consideration where special functions can be designed for special cases. Here we are trying to reach a general approach by utilizing symmetry properties and take the simplest case of optically driving a two-level atom on time interval $[-T/2, T/2]$ as an example, 
\begin{equation}
\label{eq:2level}
i\frac{d}{dt} 
\begin{bmatrix} X \\ Y \end{bmatrix}
= 
\begin{bmatrix} 0 & \frac{\Omega(t)}{2}\\ \frac{\Omega(t)}{2} & \Delta(t) \end{bmatrix}
\cdot \begin{bmatrix} X \\ Y \end{bmatrix},
\end{equation}
where $\Omega(t), \Delta(t)$ are real-valued symmetric functions. 

Next, apply the PT transformation, including the Time-reversal of letting $t \to -t$ and the Parity-inversion of letting $Y \to -Y$. The typical dipole transition occurs between two states with opposite parity and without loss of generality the postulation here is that $X$ has even parity and $Y$ has odd parity. Then we obtain the equations for the time evolution of $(X(-t), -Y(-t))$:
\begin{equation}
i\frac{d}{dt} 
\begin{bmatrix} X(-t) \\ -Y(-t) \end{bmatrix}
= 
\begin{bmatrix} 0 & \frac{\Omega(t)}{2}\\ \frac{\Omega(t)}{2} & -\Delta(t) \end{bmatrix}
\cdot \begin{bmatrix} X(-t) \\ -Y(-t) \end{bmatrix}.
\end{equation}

Now, assume that the wave function $(X, Y)$ describe some part of gate process where $X, Y$ correspond to ground and Rydberg states respectively. This study focuses mainly on the phase gate, say the CZ gate without loss of generality. Then $X(0)=1, Y(0)=0$ at the beginning and $X(T)=\pm 1, Y(T)=0$ in the end. Under the prescribed conditions, if $(X(t), Y(t))$ is a set of reasonable solution, so is $(X(-t), -Y(-t))$ with a possible overall sign change. In other words, if some appropriate waveforms $\Omega(t), \Delta(t)$, which are symmetric in time, fit for the CZ gate requirement, so do $\Omega(t), -\Delta(t)$. The above observation extends naturally to the case of buffer-atom-mediate gates involving more states and more transitions. 

The previous efforts in constructing Rydberg blockade gates with synthetic modulated driving have intentionally pursued the waveforms which are symmetric in time \cite{Yuan2024SCPMA} and explained the concepts of both off-resonant \cite{PhysRevApplied.13.024059} and resonant drivings \cite{Yuan2024FR2}. These preparations provide necessary groundwork to instantiate the abstract concepts in the above derivations and outlines the procedures to practically compute necessary waveforms. We keep the same notation of Fourier series generated waveforms, namely, the expression of waveform function $f$ is written as $[a_0, a_1, \ldots, a_N]$, representing $f(t)=2\pi \times \big(a_0 + \sum_{n=1}^{N} a_n\exp(2\pi i nt/\tau) + a^*_n\exp(-2\pi i nt/\tau) \big)/(2N+1) \text{ MHz}$ for a given reference time $\tau = 0.25\, \mu\text{s}$ which is also the gate time throughout the text. Suppression of higher frequency components and smooth on-off process are also implicitly included \cite{PhysRevApplied.20.L061002}.

\begin{figure}[h]
\centering
\includegraphics[width=0.46\textwidth]{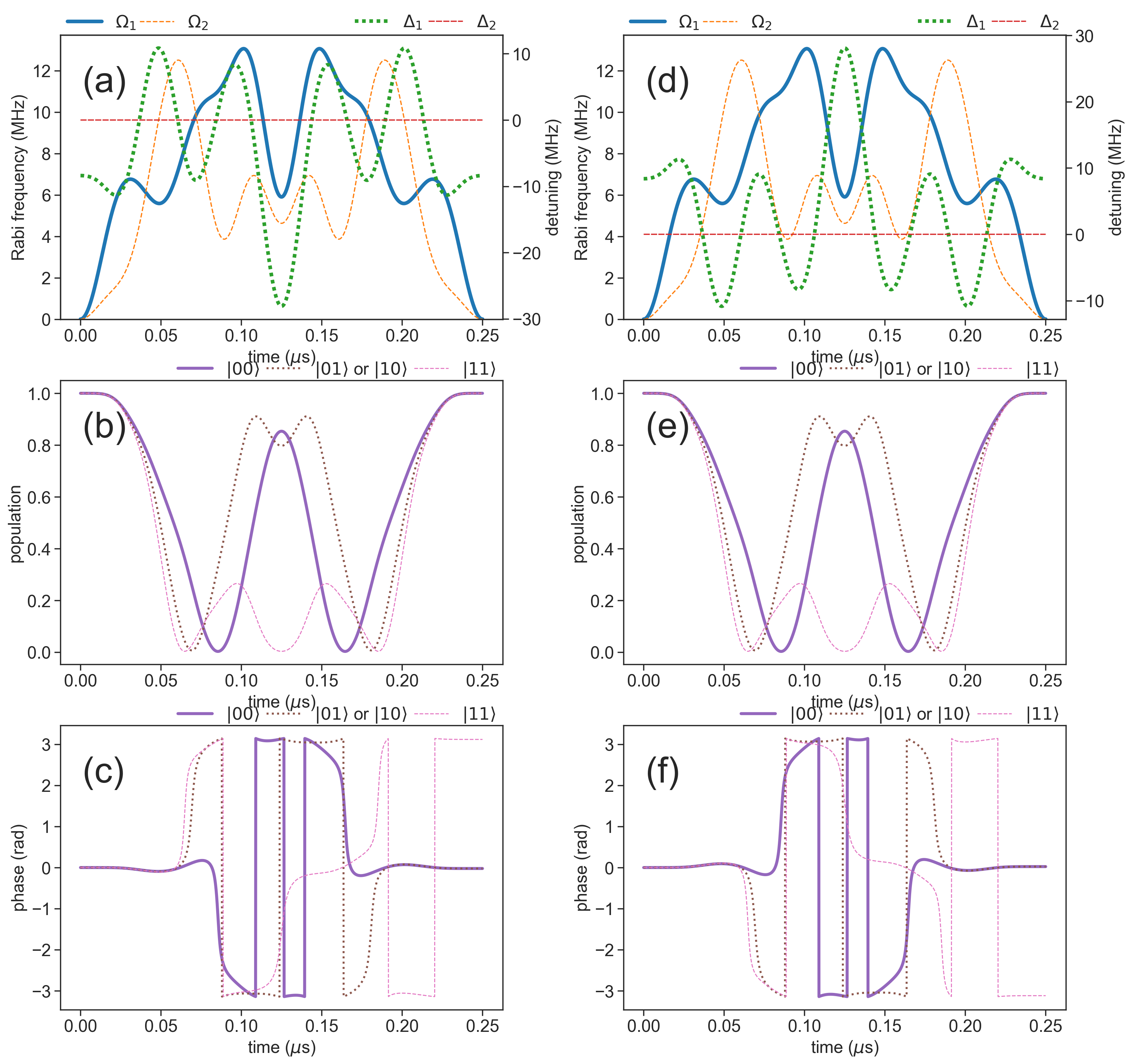}
\caption{(Color online) Sample waveforms of BAM gate with one buffer atom. (a) Waveforms of modulation.  (b) Populations of wave functions corresponding to (a). (c) Phases of wave functions corresponding to (a). (d) (e) (f) show comparisons with (a) (b) (c) after taking the PT transformation.}
\label{fig:BAMgatePT_1photon}
\end{figure}

Embedding previously discussed symmetry under PT transformation into the dual-rail technique requires the two sets of waveforms have exactly opposite detuning terms and this mandates that $\Delta_2$=0, namely off-resonant driving for the buffer atom and resonant driving for the qubit atoms. After some efforts of the numerical search in relevant parameters space, it turns out that practical solutions are attainable. Fig. \ref{fig:BAMgatePT_1photon} demonstrates the ideas here with both amplitude and frequency modulations, where both Fig. \ref{fig:BAMgatePT_1photon}(a) and its PT-reversal version Fig. \ref{fig:BAMgatePT_1photon}(d) establish CZ gate between the two qubit atoms. We consider a simplified scenario of BAM gate, where the assumptions of the Rydberg dipole-dipole interactions include $B=2\pi\times 100$ MHz for buffer-qubit pair and no interaction between two qubit atoms. The waveforms of Fig. \ref{fig:BAMgatePT_1photon}(a) consist of $\Omega_1$=[129.82, -33.36, -11.16, 5.33, -17.69, -1.62, -8.79, 4.57, -2.18], $\Omega_2$=[97.16, -16.78, -33.32, -11.58, 15.95, 5.78, -9.72, 2.92, -1.82], $\Delta_1$=[-66.80, 3.86, -63.90, 6.18, -46.78, 79.58, -2.47, -5.85, -8.46] and $\Delta_2$=0.

Very often the two-qubit gate requires coherent transition to states outside the qubit register states space. Therefore, non-negligible chances exist that some population will not return to the qubit register states space in ground-Rydberg transition due to inherent physical complexity and experimental imperfections. These population gets lost either in the Rydberg levels or intermediate levels of two-photon transition, and eventually dissipates away through the atoms' spontaneous emission. Whilst the heralded gate can possibly become helpful in dealing with gate errors associated with spontaneous emissions, we choose to omit relevant discussions in order to focus on the main theme of this paper. In other words, we can regard the main results of this paper as based on the post-selection of spontaneous emissions do not occur.

The performances of BAM gates have been extensively studied previously \cite{Yuan2024SCPMA, Yuan2024FR2} whose fidelity can hopefully reach the level of $\sim 0.999$ based on estimations with currently available state-of-the-art experiment technologies. We will omit the tedious and redundant discussions about basic properties of BAM gate itself and proceed directly to investigate the properties of heralded gates with the newly equipped features. Given the atomic properties, the key factors in the design of Rydberg blockade gates include Rabi frequencies, detunings and Rydberg blockade strengths. Adverse effects usually take place in terms of deviations from the idealized values of these terms experimentally. To analyze the properties of the heralded gate, we choose to examine overall ratio change of Rabi frequencies to check performance against Rabi frequency errors, and dc shift of detuning values to check performance against detuning errors. 

\begin{figure}[b!]
\centering
\includegraphics[width=0.46\textwidth]{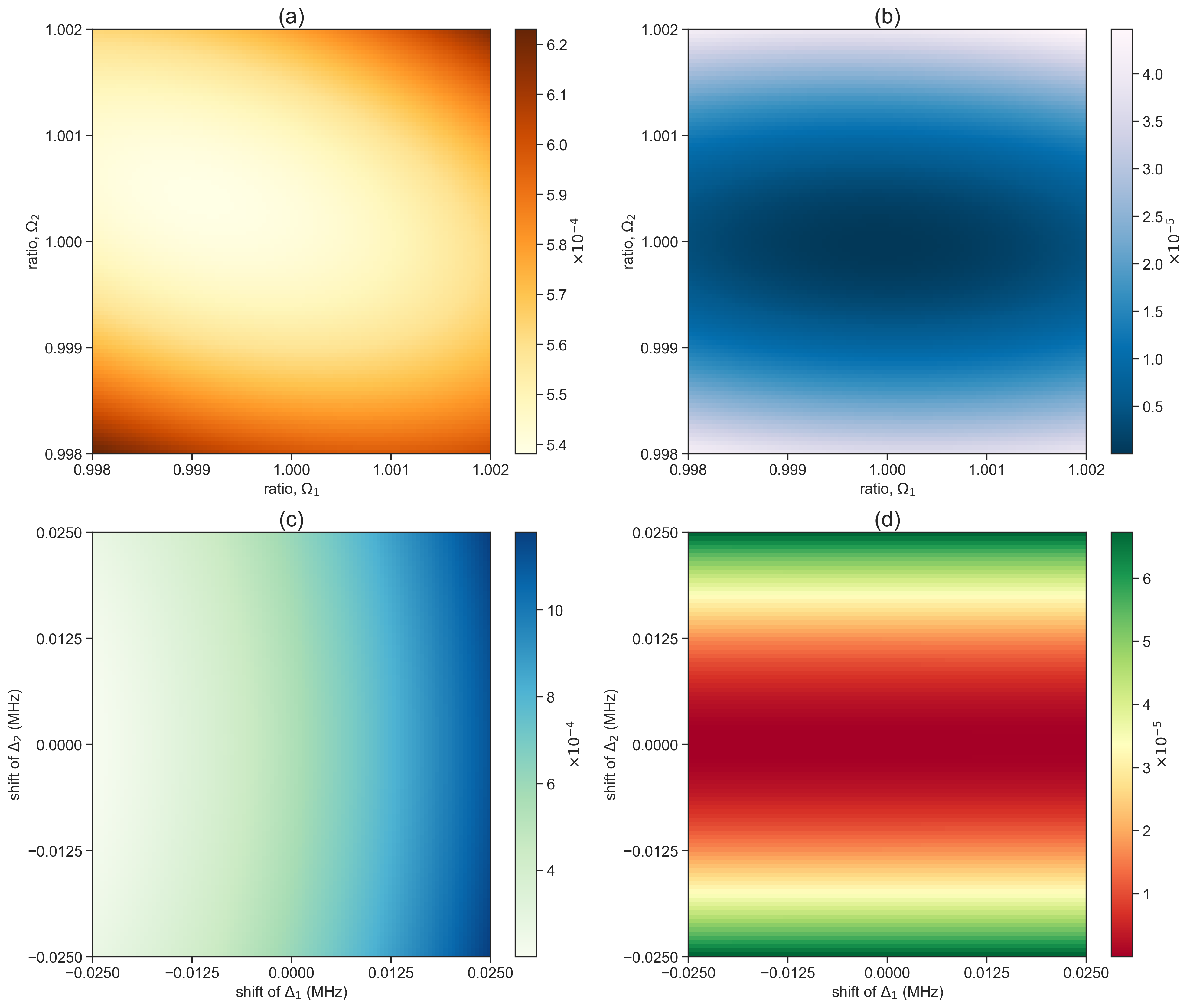}
\caption{(Color online) Numerical simulation of the heralded gate performances with respect to errors in Rabi frequencies and detunings.}
\label{fig:gate_performances_0}
\end{figure}

Fig. \ref{fig:gate_performances_0} shows a sample result of the estimated performances of waveforms in Fig. \ref{fig:BAMgatePT_1photon}, quantitatively showing the anticipated improvement of gate fidelity against Rabi frequency and detuning errors. The graph on the left shows the gate error without the heralding information. Namely, it reveals the fidelity of arriving at $|0_\text{b}\rangle\big(C_{00}|00\rangle + C_{01}|01\rangle + C_{10}|10\rangle - C_{11}|11\rangle\big)$ in step \RN{3}. On the other hand, this fidelity value also approximately behaves as the average probability of obtaining the successful heralding signal. Without the measurement outcome in step \RN{3} as the heralding knowledge, the entire system works essentially like the typical BAM gate. The graph on the right shows the fidelity of gate conditioned on obtaining the positive heralding signal. As detailed in Fig. \ref{fig:BAMgatePT_1photon}(c) and Fig. \ref{fig:BAMgatePT_1photon}(d), one needs a small local phase rotation on the buffer atom to acquire the standard phase pattern for step \RN{3} of the dual-rail technique. This sample result intentionally avoids considering or imposing this kind of local phase rotation in the calculations, in order to reveal the features of dual-rail technique. In fact, one can interpret this method as it can successfully work without knowing the exact value or even the existence of such local phase rotation, thanks to the properties of heralded gate and the special form of symmetry under PT transformation in this process. In Fig. \ref{fig:gate_performances_0} (c) \& (d), the shift $\delta$ of $\Delta_1$ means $-\Delta_1\to-\Delta_1-\delta$ for $\Omega_0$, $\Delta_1\to\Delta_1+\delta$ for $\Omega_1$ with the buffer atom and the shift $\delta$ of $\Delta_2$ means $\Delta_2\to\Delta_2+\delta$ with both qubit atoms. The absence of the local phase rotation generates the the obvious asymmetry of Fig. \ref{fig:gate_performances_0}(c), where the shift of $\Delta_1$ in the negative direction effectively compensates for that phase.

\begin{figure}[h]
\centering
\includegraphics[width=0.46\textwidth]{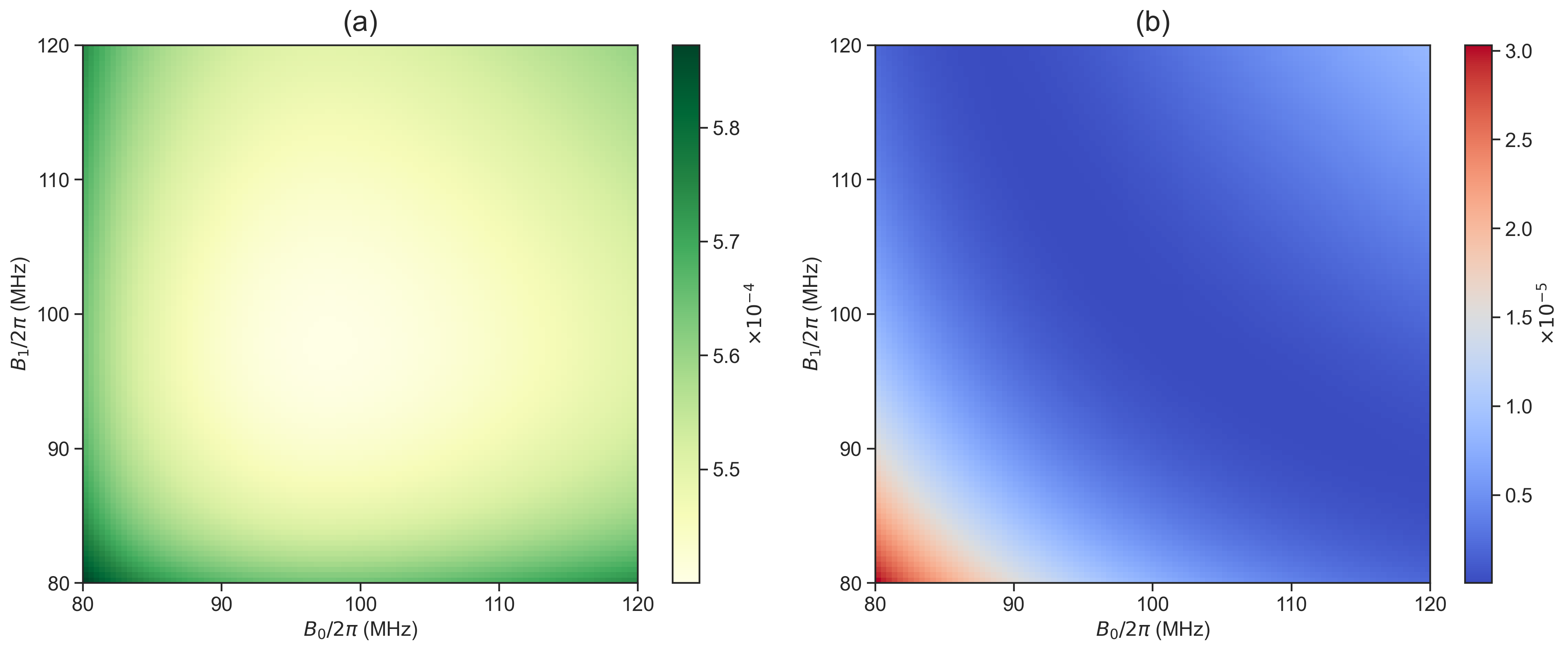}
\caption{(Color online) Numerical simulation of the heralded gate performances with respect to changes of Rydberg blockade strengths. $B_0, B_1$ stand for the strengths of two separate Rydberg dipole-dipole interaction channels respectively as given in Fig. \ref{fig:layout_sketch}(a).}
\label{fig:gate_performances_0}
\end{figure}

Next, we examine the errors caused by the change of finite Rydberg blockade strengths and the result of Fig. \ref{fig:gate_performances_0} corroborates with expectations. First of all, it comes as no surprise that the heralded gate has a higher fidelity performance even with respect to the adverse effects of varying the Rydberg dipole-dipole interaction strength. Secondly, for the dual-rail technique to behave more effectively, it is preferable for the Rydberg dipole-dipole interaction strengths to stay within certain region. At this moment, these calculations depend on the exact model of Rydberg dipole-dipole interaction and the practical condition contains significantly more complexities. Nevertheless, it is reasonable to anticipate that the main result here still holds well despite of the more complicated Rydberg levels and interaction channels by considerations from first-order perturbation \cite{OptEx480513}.

The projection of buffer atom plays a crucial role because if no measurement is made then no extra information is provided, and such that the readout fidelity needs to stay at a reasonable level. Nevertheless, these exist some subtle issues beyond a single value of readout fidelity itself. For instance, an error on the order of $10^{-4}$ does not seem to matter much for the heralding probability but definitely causes considerable trouble for the heralded gate fidelity. More specifically, the readout of buffer atom should suppress the error rate of false heralding that claims the buffer atom in state $|0_\text{b}\rangle$ while it actually does not.

\begin{figure}[b!]
\centering
\includegraphics[width=0.46\textwidth]{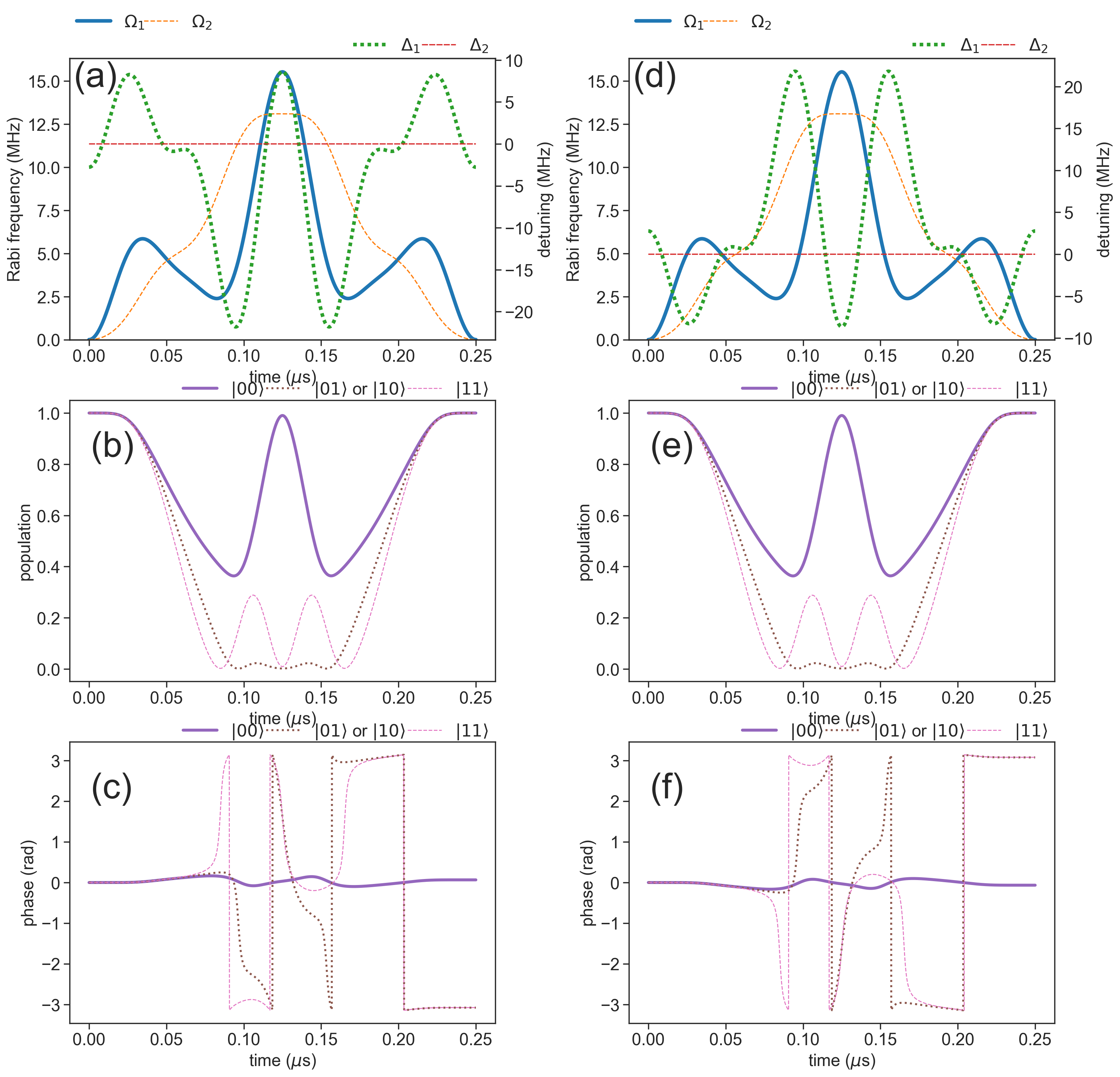}
\caption{(Color online) Sample waveforms of BAM gate under the assumption of ideal Rydberg blockade between all atoms. (a) Waveforms of modulation.  (b) Populations of wave functions corresponding to (a). (c) Phases of wave functions corresponding to (a). (d) (e) (f) show comparisons with (a) (b) (c) after taking the PT transformation.}
\label{fig:AllgatePT_1photon}
\end{figure}

The quintessence of self-correction mechanism in the heralded gates derives from the Rydberg dipole-dipole interactions between the buffer and qubit atoms. Perceivably, the buffer atom probes and compares the `two versions' of qubit atoms' wave functions and behaves like a fair judge. Fig. \ref{fig:BAMgatePT_1photon} assumes no interaction between the two qubit atoms and now let's investigate the other case where strong Rydberg blockade effect exists for all mutual pairs in this three-body system. The geometric configuration of the one buffer and two qubit atoms can look like a triangle in this case. It turns out that reasonable solutions are also attainable for this case and Fig. \ref{fig:AllgatePT_1photon} presents such a set of waveforms, with $\Omega_1$=[57.46, -17.45, 11.37, -19.97, 2.62, -5.30], $\Omega_2$=[69.00, -34.79, 3.38, -3.22, -1.85, 1.98], $\Delta_1$=[-33.12, 40.94, 9.90, -41.01, 22.85, -31.37] and $\Delta_2$=0.

\begin{figure}[h]
\centering
\includegraphics[width=0.46\textwidth]{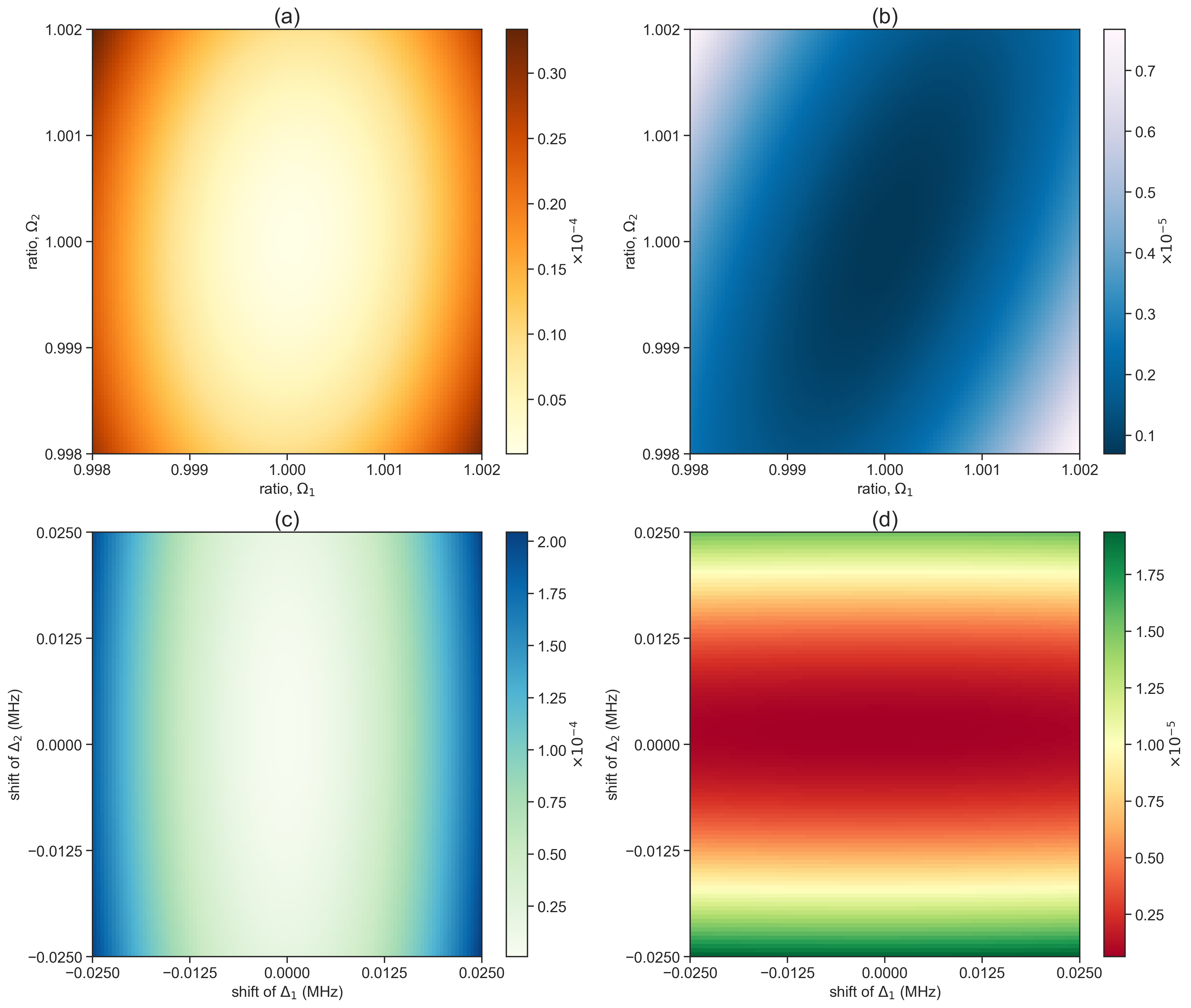}
\caption{(Color online) Numerical simulation of the heralded gate performances with respect to errors in Rabi frequencies and detunings, corresponding to the waveforms of Fig. \ref{fig:AllgatePT_1photon} under the assumption of ideal Rydberg blockade between all atoms.}
\label{fig:gate_performances_1}
\end{figure}

The estimated performances of waveforms in Fig. \ref{fig:AllgatePT_1photon} against Rabi frequency and detuning errors are presented in Fig. \ref{fig:gate_performances_1}. Here the necessary local phase rotation to arrive at the standard phase pattern of CZ gate is included in the calculations. The anticipated improvement persists in this case with different situations of mutual interactions between the atoms.

Furthermore, there exist other variation forms of the dual-rail technique to realize the heralded higher fidelity gate. For instance, Fig. \ref{fig:layout_sketch}(c) demonstrates a specific form of pulse sequence where pulse batch 1 and pulse batch 2 have the same Rabi frequencies but opposite detunings in order to satisfy the symmetry requirement. In particular, we can directly apply the symmetric Rabi frequency and detuning waveforms of BAM gates with off-resonant drivings \cite{Yuan2024SCPMA} for this purpose. The simultaneous version and pulse sequence version as outlined by Fig. \ref{fig:layout_sketch}(a) and Fig. \ref{fig:layout_sketch}(c) have subtle differences on the types of errors or deviations that they can suppress. For example, in Fig. \ref{fig:layout_sketch}(c) if all of sudden the pulse batch 2's Rabi frequencies deviate from that of pulse batch 1 by 0.1\%, then the fidelity will receive a sizable downgrade in contrast with Fig. \ref{fig:layout_sketch}(a) where such situation does not occur. On the basis of self-correction principles within these possible forms of dual-rail technique, the adverse effects undermining the fidelity of two-qubit entangling gates belong to an inherent dichotomy. More specifically, within the framework of first-order perturbation, there exist two categories of adverse effects. The first category affects only the heralding probability or the raw fidelity of BAM gates but has little influence on the heralded gate fidelity, while the other category affects both the heralding probability and heralded gate fidelity.

When adverse effects intervenes with the gate, the wave functions of qubit atoms receive not only phase deviation but also population deviation, namely the residual population in the Rydberg level not returning to qubit register states. So far we have harnessed the dual-rail technique and heralding mechanism to tackle the issue of phase deviation. Now we discuss one more patch to further enhance the gate performance against the residual population in the Rydberg level to the first order. Nevertheless, the probability of having both qubit two BAM CZ gates' qubit atoms with population in the Rydberg level belongs to higher order effects, beyond the scope of this paper. The basic concept of this patch aims at generating $\pi$ phase shift between the two wave functions with population in the Rydberg level, associated with $|0_\text{b}\rangle, |1_\text{b}\rangle$ respectively in the dual-rail process. For example, with respect to the case of pulse sequence variation form as described in Fig. \ref{fig:layout_sketch}(c), immediately after the stage of dual-pulse ground-Rydberg synthetic modulated driving of step \RN{2}, apply additional drivings to dress the qubit atoms' Rydberg state $|r\rangle$ with appropriate control such that a total phase shift of $\pi$ is accumulated for $|1_\text{c}r_\text{t}\rangle$ and $|r_\text{c}1_\text{t}\rangle$, by microwave or some form of Raman laser. Technically, with constant Rabi frequency $\Omega_r$, detuning $\Delta_r$ and pulse time $T_r$ coupling the qubit atoms' Rydberg state with some Rydberg state, the requirement can be as simple as $\sqrt{\Omega_r^2 + \Delta_r^2} \cdot T_r = 4\pi$ and $\Delta_r \cdot T_r = 2\pi$. Then in step step \RN{3} where the projection of buffer atom is made, then a successful heralding signal rules out the possibility that one of two BAM CZ gates' qubit atoms have residual population in the Rydberg level.

In conclusion, the heralded higher-fidelity two-qubit entangling gate is designed and analyzed, and the Rydberg blockade effect and BAM gate framework of neutral atom quantum computing lays down a convenient and efficient foundation for this progress. It relies on a special form of symmetry under PT transformation in the time evolution process of prescribed quantum gates to yield a self-correction behavior for deviations caused by some common sources of adverse effects, and in particular two Rydberg channels are employed to render a dual-rail version of BAM gates. The outcome wave function consists the linear superposition of higher-fidelity and lower-fidelity parts and a projection onto qubit register basis states will lead to the heralding signal. There exists a relatively high probability of getting the heralding signal which indicates the success of obtaining higher-fidelity gate. We anticipate this category of heralded Rydberg blockade gates, although probabilistic in its nature, can reduce the gate error to the level of $10^{-4}$--$10^{-6}$ or even lower which almost feels like hardware-level error correction or error mitigation, conditioned or post-selected on that the spontaneous emissions do not happen. With the fidelity of Rydberg blockade entangling gates climbing up to the next level, we are looking forward to vast new opportunities in the research of quantum computing with neutral atom qubits. We expect the results of this paper can also become useful for quantum precision measurement \cite{Kaufman2022nphys, Browaeys2023nature} and quantum sensing \cite{RevModPhys.89.035002}. Although the Rydberg blockade gates of neutral atom quantum computing bears considerable differences with the superconducting or ion qubits from the physical point of view, we are looking forward to possible variation forms of heralded higher-fidelity gate via self-error-correction process in the superconducting or ion qubit platform optimistically.

The author gratefully acknowledges supports from the Science and Technology Commission of Shanghai Municipality, the National Key R\&D Program of China, and the National Natural Science Foundation of China. The author thanks Prof. Ning Chen for many discussions about symmetry \cite{yang2003thematic}.

\bibliographystyle{apsrev4-2}

\renewcommand{\baselinestretch}{1}
\normalsize

\bibliography{trichord_ref}
\end{document}